\documentclass[%
 reprint,
 amsmath,amssymb,
 aps,
prx,
floatfix,
]{revtex4-2}

\preprint{APS-QED123}

\usepackage{graphicx}% Include figure files
\usepackage{dcolumn}% Align table columns on decimal point
\usepackage{bm}% bold math 
\usepackage{natbib}
\usepackage{appendix}
\usepackage[english]{babel}
\begin{document}

\title{Scaling in Decaying Turbulence at High Reynolds Numbers}% Force line breaks with \\

\author{Christian K\"uchler}
 \email{christian.kuechler@ds.mpg.de}
 \affiliation{Max Planck Institute for Dynamics and Self-Organization, G\"ottingen, Germany \\
 University of G\"ottingen, G\"ottingen, Germany}%Lines break automatically or can be forced with \\

\author{Eberhard Bodenschatz}
 \email{eberhard.bodenschatz@ds.mpg.de}
 \affiliation{Max Planck Institute for Dynamics and Self-Organization, G\"ottingen, Germany \\
 Physics Department,  Cornell University, Ithaca, NY, USA \\
 Institute for Dynamcis of Complex Systems, University of G\"ottingen, G\"ottingen, Germany}
\author{Gregory P. Bewley}%
 \email{gpb1@cornell.edu}    
\affiliation{%
Cornell University, Ithaca, NY, USA
}%
\date{\today}% It is always \today, today,
             %  but any date may be explicitly specified

\begin{abstract}
The way the increment statistics of turbulent velocity fluctuations scale with the increment size is a centerpiece of turbulence theories. We report data on decaying turbulence in the Max Planck Variable Density Turbulence Tunnel (VDTT), which show an approach of the inertial range statistics toward a nontrivial shape at small scales. By correcting for the contributions of energy decay to the large-scale statistics with a model, we find the scaling exponent of the second-order velocity increment statistics to be independent of the Reynolds number and equal to $0.693\pm0.003$ for $2000\lesssim R_{\lambda} \lesssim 6000$. This is evidence of a universal inertial range at high Reynolds numbers.
\end{abstract}

\pacs{Valid PACS appear here}% PACS, the Physics and Astronomy
                             % Classification Scheme.
%\keywords{Suggested keywords}%Use showkeys class option if keyword
                              %display desired
\maketitle

%\tableofcontents
\section{Introduction}\label{sec:Intro}
Turbulent fluid motion is multi-scale in space and time and its statistical properties are thought to be universal in the range of scales where the flow dynamics are governed by inertia (kinetic energy). In the inertial range  the $n$-th order moments of the velocity increments $\Delta u(r)=u(x+r)-u(x)$ , {\it i.e.} $S_n(r)=\langle (\Delta u(r))^n\rangle$, are expected to  follow scaling laws $r^{\zeta_n}$. Dimensional analysis gives $S_n\sim r^{n/3}$\cite{Kolmogorov1941}. Laboratory experiments \cite{Saddoughi1994,Kahalerras1998,Ferchichi2000,Mydlarski1996,She2017,Rousset2014,Grant1962}, field measurements \cite{Gibson1962,Praskovsky1994,Sreenivasan1998a,Gulitski2007,Tsuji2004a}, and numerical simulations \cite{Ishihara2016,Ishihara2009,Kaneda2003} support the existence of such a scaling law, but the scaling exponents $\zeta_n$ deviate substantially from $n/3$ and are a convex function of the order $n$. Much work has been invested in predicting $\zeta_n$ \cite{Kolmogorov1961,Frisch1978,Benzi1984,Sreenivasan1986, Meneveau1987,Andrews1989,Sreenivasan1991,Kida1991,She1994,Dubrulle1994,Grossmann1994,Sreenivasan1997,Mandelbrot2001} and testing them against measurements.

In fully developed statistically isotropic turbulence of an incompressible fluid, the energy is transferred predominantly from the largest spatial scales, i.e. the energy injection scale $L$, to the smallest scales, also known as the Kolmogorov scale $\eta$ with the power per unit mass $\varepsilon$. The Reynolds number quantifies the relative importance of inertial forces over viscous forces, and thus the extent of the inertial range. The Reynolds number $R_{\lambda}=u_{RMS}\lambda/\nu$ used here is based on the Taylor scale $\lambda$ \cite{Taylor1936}. 

One key question since the seminal work by Kolmogorov in 1941 \cite{Kolmogorov1941} is the extent to which $S_n$ follows power laws in the inertial range  as function of the Reynolds number. This requires high-precision measurements at high Reynolds numbers under well-controlled laboratory conditions. This has challenged experimentalists since the 1930s. Reynolds numbers in atmospheric turbulence are high, but conditions are neither controllable nor stationary. Data from such experiments \cite{Tsuji2004a,Sreenivasan1998} suggests a complicated shape of $\langle (\Delta u(r))^n\rangle$, in line with recent laboratory results of extreme statistical fidelity \cite{Sinhuber2016b}. Under the idealised conditions of numerical simulations in a periodic domain with a well-controlled energy injection mechanism, $S_n\sim r^{\zeta_n}$ is a good approximation of the data in the inertial range \cite{Cao1996,Iyer2017}. Because an increase in Reynolds number is computationally costly, and the inertial range is oftentimes very close to the turbulence excitation scale, well-converged numerical studies have been limited to relatively modest Reynolds numbers. When stopping the energy injection and leaving the turbulence to decay, rigorous methods of measuring $\zeta_n$ fail to confirm the presence of power laws in numerical studies\cite{Fukayama2000,Yang2018}. This is of great relevance for the interpretation of experimental data, because laboratory flows are usually decaying. 

An insufficient separation of scales  influences the inertial range dynamics \cite{Tang2019}. This motivated special data analysis schemes \cite{Benzi1993a, Schumacher2007} to find power laws over wider ranges of scales and thus more robust estimates of $\zeta_n$. One established technique is to use the extended self-similar scaling that appears when plotting $S_n(r)$ vs $\langle |\Delta u|^3\rangle$ known as Extended Self-Similarity (ESS).

% However, deviations from this power law are reported with similar consistency. In particular, when measuring the local scaling exponent of field \cite{Tsuji2004a} and laboratory \cite{Antonia2019} data, $\zeta_n$ is a function of $r$. The notion of a single scaling exponent is thus ambiguous and the parameters of power law fits must depend on the fitting range. These problems are partially overcome by direct numerical simulations with external forcing, where $\zeta_n$ is independent of $r$ over some very limited range of scales. The artificial forcing schemes used in these simulations likely help the creation of such a scaling range. However, the simulations are in good agreement with experimental data when considering the less idealised case of decaying turbulence \cite{Fukayama2000,Yeung2018}. 
% Finally, we do not know of any study that could show an independence of the velocity increment statistics from the Reynolds number, thus viscosity or flow geometry/forcing still played a role at all scales. 
% In this letter we show for the case $n=2$ that (a) a universal form of $\langle (\Delta u(r))^n\rangle$ can be found at finite Reynolds numbers, (b) this universal form is different from a pure power law, which can be explained by the decrease of the overall kinetic energy present in the system over time. By modelling the physics of this decay, we find (c) the scaling exponent the decay shadows. 
In this article we first study how the shape of $S_2(r)$ changes when increasing the Reynolds number to values previously out of range for well-controlled laboratory experiments. In particular, we check whether $S_2(r)$ follows a power law anywhere in the inertial range. While we cannot find such a power law, we argue that the decay of the turbulent kinetic energy reshapes $S_2(r)$ across all observed length scales in a predictable way. By applying a model \cite{Yang2018} of the decay we show that a scaling exponent can be extracted from the data. We offer a strategy to extract scaling exponents of second-order statistics based on physical modelling and compare it to other purely empirical methods, such as Extended Self-Similarity (ESS) \cite{Benzi1993a}.

\section{Experiment}
The Reynolds number of the flow in the Variable Density Turbulence Tunnel\cite{Bodenschatz2014a} (VDTT) can be finely adjusted in three largely independent ways up to levels typical for atmospheric turbulence: (i) the large-scale forcing with a novel active grid, (ii) the mean flow speed $U$ up to 5.5 m/s by adjusting the rotation frequency of its fan, and (iii) the kinematic viscosity $\nu$ by changing the static pressure. The VDTT is filled with sulfur-hexaflouride (SF$_6$) at pressures 1mbar $< p <$ 15 bar \cite{Bodenschatz2014a}.
Flow structures of variable size are introduced using a mosaic-like arrangement of individually controllable paddles ("active grid"). It allows us to obstruct the flow on finely adjustable time- and length scales \cite{Griffin2019,Kuchler2019}. In this way we control the energy injection scale between about $0.1 \mathrm{m} \lesssim L \lesssim 0.6 \mathrm{m}$. 
The small kinematic viscosity of pressurized SF$_6$ permits the existence of very small flow structures. The size of these structures scales with the dissipation length $\eta=(\nu^3/\varepsilon)^{1/4}$, where $\varepsilon =15\nu \langle (\partial u/\partial x)^2\rangle$. For the range of ambient pressures 1 bar $< p <$ 15 bar, this dissipation length is between $250 \mathrm{\mu m} \gtrsim \eta \gtrsim 10 \mathrm{\mu m}$.

We record time series of hot-wire signals and convert them into one-dimensional flow fields assuming that the turbulent fluctuations are passively advedcted across the sensor by the mean flow $U$. Thus, a time step $\Delta t$ is converted to a spatial increment $\Delta x = U \Delta t$ \cite{Taylor1938}. We use a commercial constant temperature anemometer to drive and acquire data from Nanoscale Thermal Anemometry Probes (NSTAP) provided by Princeton University \cite{Kunkel2006,NSTAP1,Fan2015}. These ultra-small hot wire probes average the flow field over a length of only 30$\mu$m, which is sufficient for this experiment. For flows where the turbulence length scales are larger, we also use commercial hot-wires from Dantec Dynamics with sensing length 450$\mu$m ($\gtrsim 4 \eta$).
The frequencies (and wavenumbers) encountered in the measurements presented here are in a range that is not particularly demanding for this combination of sensor and anemometer circuitry \cite{Hutchins2015,Samie2018a,Ashok2012}. The experiments presented here were taken under different ambient pressures and different active grid forcing schemes to allow for a careful check of the hot wire fidelity. We thus ensure the robustness of the results against probe- or flow geometry-induced biases. We emphasise that all conclusions presented here are independent of the type of probe used (NSTAP or commercial hot wire), the frequencies where turbulent fluctuations are measured, the dissipation length scale, and the active grid forcing (see Supplementary Material for more details).

\section{Small Scale Universality \label{SmallScaleUniversality}}

\begin{figure*}
		\includegraphics[width =\textwidth]{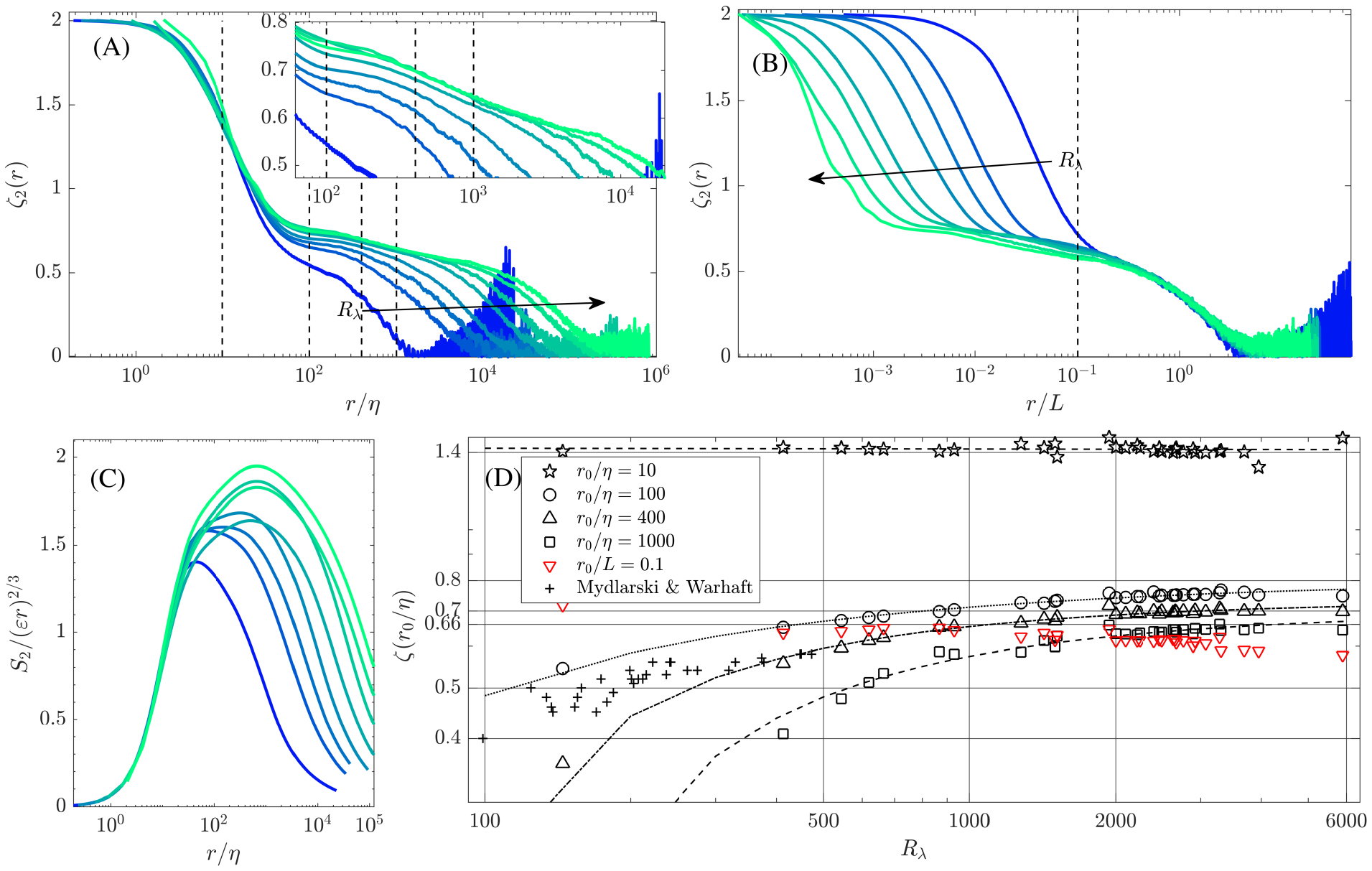}
        \caption{(A): $\zeta_2(r)$ for $R_\lambda$=144, 413, 620, 931, 1520, 2400, 3690, and 5890. The curves collapse approximately to a universal form for $R_{\lambda}>2000$ that extends from the smallest scales up to $0.1L$ and is different from a constant, which indicates that power law scaling is masked in these data. In contrast, the curves at $R_{\lambda}<2000$ change significantly with $R_{\lambda}$. Inset: Zoom on the inertial range of the same curves. At the largest $R_\lambda$ a wave-like fine structure can be seen as in \cite{Sinhuber2016b}. Dashed lines: $r_0/\eta$ of curves in (D);  (B): Same as (A), but normalised by $L$. The curves approach a universal form from the largest scales down to $0.2L$. (C): Structure functions $S_2$ compensated by the self-similar prediction $\sim (\varepsilon r)^{2/3}$. (D): $\zeta_2(r)$ evaluated at fixed $r_0/\eta$. The curves approach constants, but their values depend on $r$. Thus, the value of $\zeta_2$ (assumed constant across a wide range of scales and universal in $R_{\lambda}$ in many turbulence models) is a function of both $r$ and $R_{\lambda}$. The curves saturate at finite $R_{\lambda}$ indicating that this apparent discrepancy with the models persists as long as $R_{\lambda}$ takes finite values. The lines are fits of  $C_1-C_2 R_{\lambda}^{\alpha}$.
        The red inverted triangles correspond to the dashed line in (B), i.e. they show the scaling exponent at a fixed scale relative to $L$. Black points above this curve are within the inertial range (except for the case $r_0/\eta=10$).}
        		\label{LocalSlopeCollapseEta}

\end{figure*}

We investigate whether the shape of the velocity increment statistics  approach a universal form at large Reynolds numbers $R_{\lambda}$. In particular, we seek a universal scaling exponent $\zeta_n$ such that $S_n\sim r^{\zeta_n}$. 

A rigorous method to find and extract power laws is to calculate the local scaling exponents
\begin{equation}
  \zeta_n(r)=\frac{d \log(S_n)}{d \log(r)}.
  \label{eqn:DefLogDer}
\end{equation}
 $\zeta_n(r)$ is constant when power law scaling exists. Fig. \ref{LocalSlopeCollapseEta} (A) shows measurements of $\zeta_2(r)$ for flows at different $R_\lambda$. Above some finite $R_\lambda\approx 2000$ the curves begin to collapse from the dissipation range up to $\approx 0.1L$. The collapse in the dissipation range $r < 20\eta$ is expected and the exponent at $r\rightarrow0$ corresponds to a Taylor expansion around that point. Around $r/\eta=100$ the curves deviate slightly from each other even for $R_\lambda>2000$. This region is influenced by the bottleneck effect \cite{Sinhuber2016b,Falkovich1994,Donzis2010a,Verma2005,Lohse1995}, whose Reynolds number dependence even at high $R_\lambda$ has been shown \cite{Kuchler2019}. Above the bottleneck in the inertial range,  the curves collapse again for $R_\lambda>2000$, but they are not flat as would be expected for a power law. In particular, the self-similar scaling, where $S_2/(\varepsilon r)^{2/3}= \mathrm{const}$ cannot be found, as shown in Fig. \ref{LocalSlopeCollapseEta} (D). In general, no single scaling exponent can be discerned at all (see Fig. \ref{LocalSlopeCollapseEta} (A)-(B)). 

While the inset in Fig. \ref{LocalSlopeCollapseEta} (A) already shows that the shape of $\zeta_2(r)$ stops changing with increasing $R_\lambda$, we address this with greater rigour in Fig. \ref{LocalSlopeCollapseEta} (D). We pick several values of $r_0/\eta$ and plot $\zeta_2(r_0/\eta)$  for different $R_\lambda$. For each $r_0/\eta$, the curves approach a different constant at high $R_\lambda$. In contrast, if a scaling exponent were to emerge at even higher $R_\lambda$, these curves would approach a common constant independent of $r_0/\eta$ (as long as $r_0$ is in the inertial range). Fig. \ref{LocalSlopeCollapseEta} (B) shows that equivalent arguments can be made when normalizing by the energy injection scale $L$ instead of the dissipation scale $\eta$. 
These findings are in qualitative agreement with the observation that a large scaling region can be found when plotting $S_n(r)$ vs $\langle |\Delta u|^3 \rangle$ \cite{Benzi1993a} (ESS). While it has little physical foundation except  its empirical success, this strategy yields more robust estimates of the scaling exponent. A closer look at ESS plots reveals a fine structure that looks like an oscillation around the general trend throughout the inertial range\cite{Sinhuber2016}. Because of the high $R_\lambda$, this  fine structure is now also apparent in $\zeta_2(r)$ itself. 

\section{Flow Unsteadiness Shapes Structure Functions}
\begin{figure}
  \includegraphics[width=\columnwidth]{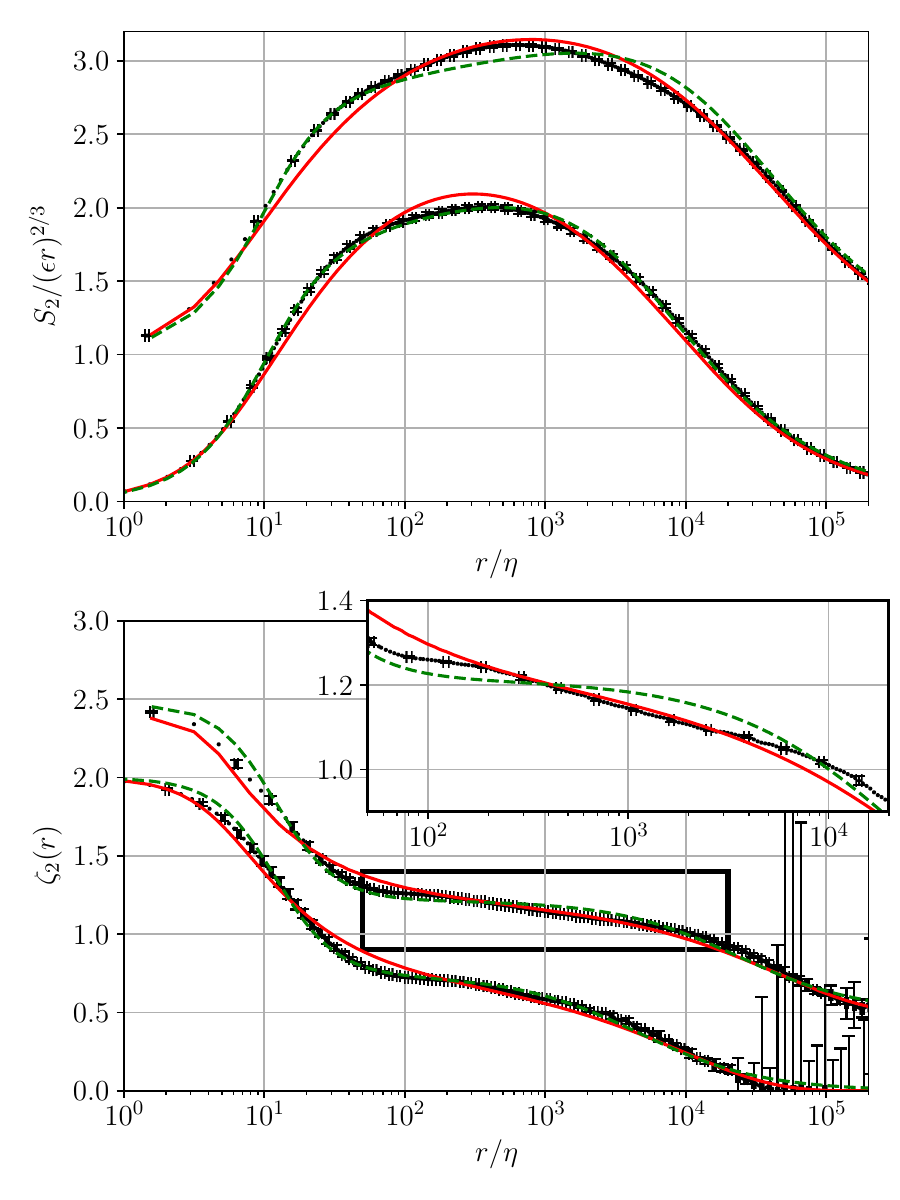}
  \caption{Demonstration of the fit results for $R_\lambda=1600$ (lower curves) and $R_\lambda = 3700$ (upper curves, offset for clarity). The black dots are experimental data, the red curves are the fit to the physics-based model eq. (\ref{eqn:YangModel}) with two fit parameters. The green curves represent four-parameter fits of the Batchelor interpolation formula. The region of interest is the inertial range between $100 < r/\eta < 10000$ in the high-$R_{\lambda}$ case as highlighted by the inset, where the red curves follow the experimental data more closely. For $r/\eta < 100$ the Batchelor interpolation is superior, where the physics-based model is expected to perform poorly. At scales larger than the inertial range, both fits have a similar quality.}
  \label{fig:Fits}
\end{figure}

In the previous section we show that the local slopes $\zeta_2(r)$ of the second order structure function $S_2(r)$ take a universal form at finite $R_{\lambda}$ different from a clear power law scaling. This was also reported in other laboratory \cite{Sinhuber2016b,Antonia2019} and field \cite{Tsuji2004a} experiments, but is in contrast to forced direct numerical integrations of the Navier-Stokes-equations\cite{Iyer2017}, where $\zeta_2(r)$ is flatter in the inertial range. The main difference between experiments and simulations is that the latter is usually forced continuously, whereas most measurements can be performed in unsteady flow states only, where the turbulent kinetic energy decays over time.  $\zeta_2(r)$ is tilted in simulations of decaying turbulence\cite{Yang2018, Fukayama2000} as it is in experiments \cite{Antonia2019}. Refs. \cite{Danaila2002,Boschung2016} provide an example of theoretical considerations regarding the effect of flow unsteadiness on the velocity increment statistics. 
In our experiment, the turbulent kinetic energy $u_{RMS}^2$ decays along the length of the measurement section, but the integral length scale $L$ remains constant or also decays over time. This is in contrast to freely decaying turbulence, where $L$ grows with time\cite{Davidson2015,Sinhuber2015}. We believe that the boundaries of the measurement section  with cross-section 1.2 m $\times$ 1.5 m (with 0.1 m $\lesssim L \lesssim $ 0.6 m) suppresses this growth. We found this to be relatively independent of the way we estimate $L$. We chose to use $L=\int_0^{r_s} \langle u(x)u(x+r) \rangle/u_{RMS}^2 dr$ with $\langle u(x)u(x+r_s) \rangle=0$. Other definitions of $\varepsilon$ impact the results at small $R_\lambda$, but do not affect the conclusions.

Recent work in Yang et al. \cite{Yang2018} provides a model spectrum based on decaying turbulence in a confined domain. It rests on the assumption that the energy spectrum of the turbulent fluctuations $E(k,t)$ consists of a small-scale term, a large-scale term and the inertial range scaling $k^{-5/3}$ (equivalent to the self-similar scaling of $S_2\sim r^{2/3}$). Most importantly, the model assumes a self-similar decay of turbulent kinetic energy, i.e. $u_{RMS}(t)\sim t^{-\alpha}$, and applies a model \cite{Pao1965} for the scale-by-scale energy transfer term in the evolution equation of the energy spectrum (see Appendix \ref{app:ModelSpec} in \cite{Yang2018} for details). 
In a confined domain, where the parameter describing $d L/d t$ tends to zero, this model predicts an energy spectrum of the form
\begin{equation}
  E(k)\sim\frac{-A_K}{C}(kL)^{-(\zeta_{2F}+1)}e^{(3A_K/C)(kL)^{-2/3}}e^{-(1.5/C)(k\eta)^{4/3}}.
  \label{eqn:YangModel}
\end{equation}

 For the purpose of measuring a scaling exponent, we replaced the term $(kL)^{-5/3}$ with $(kL)^{-(\zeta_{2F}+1)}$, where the fitting parameter $\zeta_{2F}$ is the inertial range scaling exponent for the second order structure function\cite{Pope2000}.
Because structure functions are local in space (they can be regarded as a very basic wavelet transform\cite{Farge2001}), they are smoother than the energy spectrum and easier to analyse in our data. The one-dimensional versions of $S_2$ and $E(k)$ are related through the following integral transform \cite{Monin1976}:
\begin{equation}
  S_2(r)=\int_0^{\infty}E(k) \left(\frac{1}{3}+\frac{\cos(kr)}{(kr)^2}-\frac{\sin(kr)}{(kr)^3}\right) dk.
  \label{eqn:IntegralTransform}
\end{equation}
We fit the logarithmic derivative eq.(\ref{eqn:DefLogDer}) of $S_2$ calculated from (\ref{eqn:YangModel}) to the measured values of $\zeta_2(r)$ using eq. (\ref{eqn:IntegralTransform}).
Fig.\ref{fig:Fits} compares the experimental data with the model (\ref{eqn:YangModel}) and a three-parameter Batchelor fit. The latter is is a parametric fit that models the structure function shape without physical justification\cite{Dhruva2000,Lohse1995}. The Batchelor fit appears to be slightly superior in the case of low $R_\lambda$ (lower curves). However, as $R_{\lambda}$ increases, it cannot follow the inertial range shape. In contrast, eq. (\ref{eqn:YangModel}) does follow the experimental data in the inertial range. The deviations around $r/\eta=60$ are due to the bottleneck effect described earlier.
The model (\ref{eqn:YangModel}) allows us to measure $\langle\zeta_{2F}\rangle$ and $A_k$. 

\begin{figure}
  \begin{center}
  \includegraphics[width=1\columnwidth]{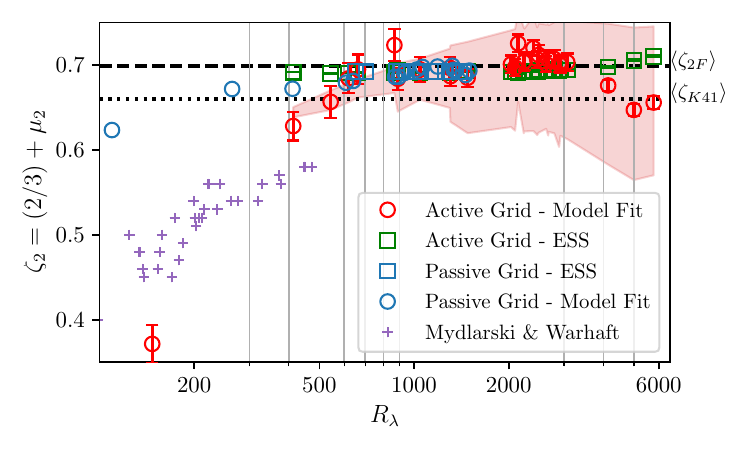}
  \caption{Circles: Results of fitting the parameters in eq. (\ref{eqn:YangModel}). $\zeta_2$ approaches a constant $\langle\zeta_{2F}\rangle=0.693\pm0.011$ (dashed), higher than Kolmogorov's prediction (dotted) \cite{Kolmogorov1941}. We attribute the slight downward trend in last two data points of $\langle\zeta_{2F}\rangle$ to the anisotropic grid forcing that was used to reach these high $R_\lambda$. For comparison we show the extended self-similarity estimates (squares), i.e. the scaling of $S_2(|S_3|)$ and the data from Mydlarski $\&$ Warhaft \cite{Mydlarski1996}. The shaded region corresponds to the values that the local slope $\zeta_2(r)$ takes within $100\eta<r<0.1L$.}
  \label{fig:FitResults}
  \end{center}
\end{figure}

We now analyse the dependence of these model parameters on the Reynolds number, in particular the inertial range scaling exponent $\zeta_{2F}$. Fig. \ref{fig:FitResults} shows how $\zeta_{2F}$ depends on $R_\lambda$. We observe the approach towards a common scaling exponent around $\langle\zeta_{2F}\rangle = 0.69$ for $R_\lambda\gtrsim 2000$. This is a little higher than the prediction for intermittency-free turbulence $\zeta_{2}=2/3$ and almost identical to the values that can be read off typical DNS data \cite{Kaneda2003,Ishihara2009,Gotoh2002,Cao1996} and the ESS estimate from the data. However, the ESS scaling exponent is practically constant around 0.69 over all $R_\lambda$. The variation of $\zeta_{2F}$ over one decade of $R_{\lambda}$ is much smaller than the variation in $\zeta_2(r)$ within a single measurement in the inertial range. Finally we compare the data to the measurements by Mydlarski $\&$ Warhaft \cite{Mydlarski1996} and indicate the range of values that $\zeta_2(r)$ takes between $r/\eta=100$ and $r=0.1L$ in our measurements (shaded region).
We finally revisit data that was acquired in the VDTT in the far-field of a passive grid and apply our analysis to extract a scaling exponent. For a passive grid the decay in this experiment has been shown by Sinhuber et al. \cite{Sinhuber2015} to follow the self-similar decay predicted by Saffman \cite{Saffman1967}. In this case, the integral scale grows as a function of the decay time. However, as can be seen from Fig. 3 in their paper \cite{Sinhuber2015}, the integral length saturates at the most downstream positions. The reason is likely the same as in the active grid case investigated here: The tunnel boundaries inhibit a further growth of $L$. To make a fair comparison, we use those downstream positions in the passive grid data and apply the theory for the case of $L=$ const. 

The parameters $C$ and $A_K$ are related through $C=-A_K(6/\pi)^{1/3}$. In practice, $A_K$ describes the large-scale part of the energy spectrum, which is heavily influenced by the decay. For formal definitions see Appendix \ref{app:ModelSpec} as well as the original publication of the model \cite{Yang2018}. In our measurements, $A_K$ depends only slightly on $R_{\lambda}$ at low Reynolds numbers, but approaches a constant $A_K\sim -0.55$ for $2000 \lesssim R_{\lambda} \lesssim 6000$.  This is consistent with Fig. \ref{LocalSlopeCollapseEta} (B) and earlier studies \cite{Sinhuber2015} that measured that the decay exponent and thus the large-scale part of the energy spectrum are largely independent of the Reynolds number. It further agrees with the observation that $A_K$ is related to the cascade efficiency $C_{\varepsilon}=\varepsilon L/u_{RMS}^3=(-A_K)^{3/2}$ \cite{Yang2018,Vassilicos2015,Sreenivasan1998b,Sreenivasan1984}.

Note a possible qualitative inconsistency between the measurements presented here and the model \cite{Yang2018} in the limit of $R_\lambda\rightarrow \infty$. For this limit, the model assumes the approach to a constant $\zeta_2(r)$ valid for the entire inertial range. Our measurements (see Fig. \ref{LocalSlopeCollapseEta} (A) and (D)) suggest that a further increase in $R_\lambda$ does not lead to such a power law. However, final conclusions about this limit cannot be made on the basis of our data and measurements at higher $R_{lambda}$ than presented here would be needed.

\section{Discussion}
In this paper we consider the foundation of multiple models of turbulence, namely the presence of a well-defined power law scaling in the inertial range. We confirm experimental results that such a power law is a reasonable approximation to the shape of the second-order velocity increment moments $S_2(r)$. Existing differences between the experimental data, theoretical predictions, and numerical simulations were in the past attributed to an insufficient range of inertial scales \cite{Tang2019,Antonia2019}. There seemed to be a consensus that in a wind tunnel at extremely large Reynolds numbers, viscosity and flow geometry would become unimportant, and $S_n$ would approach a clear power law of exponent $\zeta_n$. We test this consensus first by analysing the logarithmic derivative of $S_2$, which is a measure of the local scaling exponent $\zeta_2(r)$. We find that $S_2(r)$ measured by $\zeta_2(r)$ follows a universal function from the dissipation scales up to the largest scales in the flow, but does not follow a power law. In particular, increasing $R_{\lambda}$ beyond 2000 does not improve the validity of $S_2\sim r^{\zeta_2}$ further. Even at $R_\lambda \approx 6000$ we cannot find a range of scales where $\zeta_2 = \mathrm{const}$ and our data suggest that further increases in $R_\lambda$ would not change this. 

In a next step, we explain these observations by considering the decay of the turbulent kinetic energy present in the flow. A model of the energy spectrum derived from a popular closure model\cite{Pao1965} while accounting for the decay\cite{Yang2018}, yields predictions of $S_2$ that are superior to those of a interpolation formula (Batchelor) aiming to empirically describe the effects of finite Reynolds numbers alone \cite{Dhruva2000}. Our explanation is also supported by observations from numerical simulations: While $\zeta_2(r) = \mathrm{const.}$ can be found in numerical simulations even at moderate Reynolds numbers \cite{Kaneda2003}, this is not possible in simulations of decaying turbulence \cite{Fukayama2000,Yang2018}.

While the model predicts an approach to a power law scaling at finite, but extremely high $R_{\lambda}$, Figs. \ref{LocalSlopeCollapseEta} (A) and (D) support a different conclusion: The flow unsteadiness (decay) reshapes the flow statistics from the largest flow scales $r \approx L$ throughout the entire inertial range. We have no experimental evidence that this influence of the decay vanishes at even higher $R_{\lambda}$ and measurements in the atmospheric surface layer \cite{Tsuji2004a,Sreenivasan1998,Dhruva1997} seem to support this up to $R_{\lambda}<20000$. Finally, the closure theory \cite{Pao1965}, on which the model spectrum is based, is known to fail at dissipative scales. 

We used knowledge about the influence of the decay on the inertial range to extract an inertial range scaling exponent for $S_2$ with the  assumption that $S_2(r)$ follows a power law shadowed by the effects of decay. We arrive at a value value of $\langle \zeta_{2F} \rangle = 0.693 \pm 0.003$ by averaging all values of $\zeta_{2F}$ measured at $R_{\lambda}>2000$. This value is extremely close to the value $\langle \zeta_{ESS} \rangle = 0.692 \pm 0.001$ extracted by comparing structure functions of different order (extended self-similarity, ESS). This suggests that the physical processes underlying the almost 30-year long successful application of ESS are small-scale universality combined with large-scale effects influencing almost all turbulence length scales. 

The scaling exponent $\langle \zeta_{2F} \rangle$ extracted from the model fit shows a clear $R_\lambda$-dependence, whereas the ESS estimate is constant over the whole range of $R_\lambda$. This study indicates that the second-order statistics scale differently at small and high $R_{\lambda}$. Nevertheless, the asymptotic scaling exponents we extract agree with the ESS estimates at much lower Reynolds numbers. It is therefore a matter of future studies to elucidate the underlying reasons. 
Finally, recent experimental results\cite{Sinhuber2016b} suggest that dissipative effects occur over the entire inertial range, which is in agreement with small-scale universality from the smallest scales up to $0.1L$ observed here.  We can observe these effects directly in $\zeta_2(r)$ in the form of oscillations around the general trend due to the extremely long inertial range. 

By creating an inertial range with unmatched control over the flow parameters, we show that the inertial range is influenced in detail by both dissipation and decay for all observed Reynolds numbers. Most importantly, we show that $S_2$ follows the same shape at small scales for all $R_{\lambda}>2000$. 

The next steps include finding functional forms that describe $\zeta_n(r)$, to extract scaling exponents of arbitrary order, and compare them to existing theories for $\zeta_n$. 

\section{Acknowledgements}
We thank Markus Hultmark and Yuyang Fan for providing the nanoscale hot wire probes and helping with their operation. We thank M. Sinhuber for help with using the passive grid data and helpful discussions. We thank A. Pumir, H. Xu, M. Wilczek, and D. Lohse for helpful discussions. The VDTT is maintained and operated by A. Kubitzek, A. Kopp, and A. Renner. The machine workshop led by U. Schminke and the electronic workshop led by O. Kurre built and installed the active grid. The Max Planck Society and Volkswagen Foundation provided financial support for building the VDTT. 

\bibliography{references1.bib}

\appendix
\section{Measurements of Decay}
\begin{figure}
    \centering
    \includegraphics[width=\columnwidth]{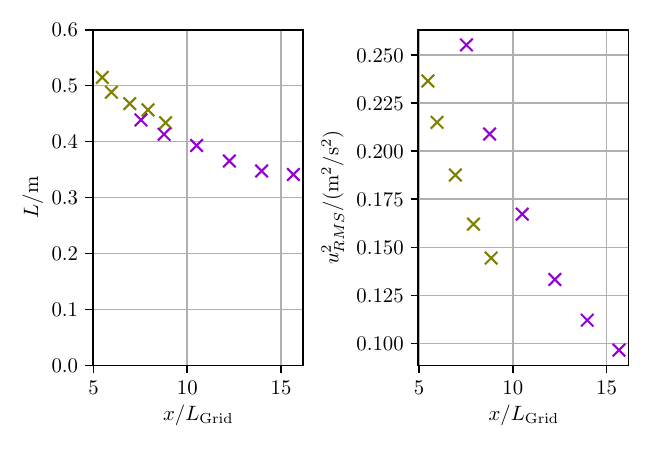}
    \caption{Development of the integral scale (left) and turbulent kinetic energy (right) for different distances from the grid. The distances are normalised by the active grid length scale defined by the correlation lengths of the paddle protocol (see \cite{Kuchler2019} for details). This scale is different from the mesh size normally used in passive grid turbulence. $L$ was estimated using from $L=\int_0^{r_s} \langle u(x)u(x+r) \rangle/u_{RMS}^2 dr$, results for different definitions of $L$ are similar.}    
    \label{fig:DecayPlot}
\end{figure}

We have measured time series of velocity fluctuations along the centerline of the measurement section in various distances from the active grid with two different forcing mechanisms that produce different energy injection scales. Fig \ref{fig:DecayPlot} shows that the turbulent kinetic energy measured by $u_{RMS}^2$ is decreasing at all points considered here for both grid protocols. It further shows that the integral scale is not growing, but slightly decreasing. 

\section{Experimental Conditions} 
\begin{figure}[htb]
    \centering
    \includegraphics[width=\columnwidth]{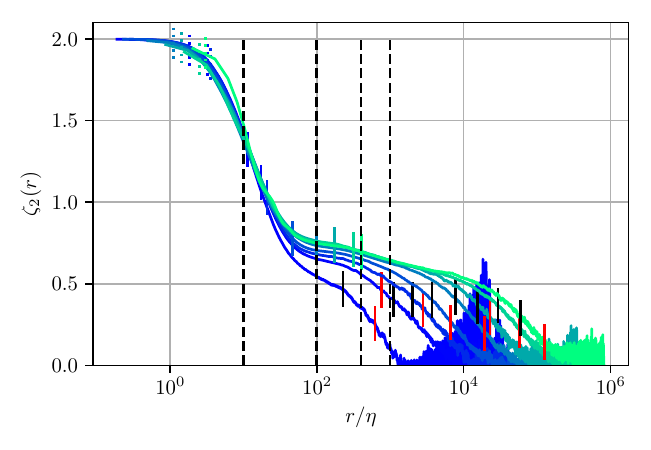}
    \caption{Same as Fig. \ref{LocalSlopeCollapseEta} (A) with the addition of probe length (dotted vertical lines), the value of $r/\eta$ corresponding to a measurement frequency of 1 kHz through Taylor's Hypothesis (vertical lines), the values of $r_0/\eta$ chosen to assemble Fig. \ref{LocalSlopeCollapseEta} (D) (dashed black lines), the length of the energy injection scale (vertical black lines), and the grid length scale (red lines). }
    \label{fig:ExperimentalConditions}
\end{figure}

Fig \ref{fig:ExperimentalConditions} indicates critical experimental length scales along the measurements of $\zeta_2$. The probe averaging length mainly influences smaller scales and is far away from the region of interest. The temporal resolution is determined by the noise filtering frequency and the frequency response of the measurement system. The frequency response of the system is not perfectly flat anymore starting around 1kHz \cite{Hutchins2015}. The range of scales we are interested in is therefore in the flat part of the frequency response curve. The noise filtering frequency is always at frequencies above 1kHz. 

\section{The Model Spectrum}\label{app:ModelSpec}
The evolution equation of the energy spectrum $E(k,t)$ can be derived directly from the Navier-Stokes-Equation and is known as the Karman-Howarth-Lin equation.
\begin{equation}
  \partial_t E(k,t) = -\partial_k\Pi(k,t)-2\nu k^2E(k,t).
  \label{KarmanHowarth}
\end{equation}
The first term on the RHS describes the nonlinear transfer of energy from small to large wavenumbers and ultimately prevents the closure of the equation. The closure approach used in the model by Yang et al. was first suggested by Pao and assumes that the transfer term $\Pi$ is local in wavenumber space and has a self-similar form:
\begin{equation}
    \Pi(k,t)=\varepsilon^{1/3}k^{5/3}E(k,t)
\end{equation}
The second term on the RHS represents the viscous dissipation at the smallest flow scales. This yields a closed form of the Karman-Howerth-Lin equation. The model further assumes that the energy spectrum can be assembled by a large scale term $f_L(kL)$ a small scale term $f_\eta(k\eta)$ and a self-similar inertial range:
\begin{equation}
    E(k,t)= C_k \varepsilon^{2/3} k^{5/3} f_\eta(k\eta) f_L(k L)
\end{equation}
To extract a scaling exponent, $k^{5/3}$ has been replaced by $k^{\zeta}$ for our purposes and $\zeta_2 = -\zeta-1$. 
It can be shown that $C=-A_K(6/\pi)^{1/3}$. This quantity is related to the dissipation constant $C_{\varepsilon}=\varepsilon L/u^3$ relating the large scale energy injection and the small scale energy transfer rate $\varepsilon$. $A_K$ is the non-dimensionalized time-evolution of the energy spectrum prefactor $d (C_K \varepsilon^{2/3})/dt$, which is a free parameter.

\end{document}